\documentclass[aps,prl,preprint,superscriptaddress]{revtex4-2}
\usepackage{braket}
\usepackage{graphicx}
\usepackage{amsmath}
\usepackage{color}

\begin{document}

\title{Excite atom-photon bound state inside the coupled-resonator waveguide coupled with a giant atom}

\author{Han Xiao}
\altaffiliation{These authors contribute equally to this work.}
\affiliation{State Key Laboratory of Advanced Optical Communication Systems and Networks,\\ School of Physics and Astronomy, Shanghai Jiao Tong University, Shanghai 200240, China}
\author{Luojia Wang}
\altaffiliation{These authors contribute equally to this work.}
\affiliation{State Key Laboratory of Advanced Optical Communication Systems and Networks,\\ School of Physics and Astronomy, Shanghai Jiao Tong University, Shanghai 200240, China}
\author{Zhenghong Li}
\email{refirefox@shu.edu.cn}
\affiliation{Department of Physics, Shanghai University, Shanghai 200444, China}
\affiliation{Zhejiang Province Key Laboratory of Quantum Technology and Device, Zhejiang University, Hangzhou 310027, China}
\author{Xianfeng Chen}
\affiliation{State Key Laboratory of Advanced Optical Communication Systems and Networks,\\ School of Physics and Astronomy, Shanghai Jiao Tong University, Shanghai 200240, China}
\affiliation{Shanghai Research Center for Quantum Sciences, Shanghai 201315, China}
\affiliation{Jinan Institute of Quantum Technology, Jinan 250101, China}
\affiliation{Collaborative Innovation Center of Light Manipulation and Applications, Shandong Normal University, Jinan 250358, China}
\author{Luqi Yuan}
\email{yuanluqi@sjtu.edu.cn}
\affiliation{State Key Laboratory of Advanced Optical Communication Systems and Networks,\\ School of Physics and Astronomy, Shanghai Jiao Tong University, Shanghai 200240, China}

\date{\today}

\begin{abstract}
It is of fundamental interest in controlling the light-matter interaction for a long time in the field of quantum information processing.
However, usual excitation with the propagating photon can hardly excite a localized state of light while keeping the atom under a subradiant decay in an atom-waveguide system.
Here, we propose a model of coupling between a giant atom and the dynamically-modulated coupled-resonator waveguide and find that a bound state, where the light shows the localization effect and atom exhibits a subradiant decay time, can be excited by a propagating photon.
An analytical treatment based on the separation of the propagating states and localized states of light has been used and provides inspiring explanation of our finding, i.e., a propagating photon can be efficiently converted to the localized light through the light-atom interactions in three resonators at frequency difference precisely equivalent to external modulation frequency.
Our work therefore provides an alternative method for actively localizing the photon in a modulated coupled-resonator waveguide system interacting with giant atom, and also points out a way to study the light-atom interaction in a synthetic frequency dimension that holds the similar Hamiltonian.
\end{abstract}

\maketitle

\newpage

It is of great importance in achieving flexible manipulations of photons in atom-waveguide systems and exploring fundamental physics associated with strong light-atom interactions and atom-mediated photon-photon interactions, which also shows potential applications towards quantum communications and quantum networks \cite{Goban2015,Pichler2016,Forn-Diaz2017,Kumlin2018,Mahmoodian2018,Corzo2019,Mirhosseini2019,Xiao2020,Kim2021}.
Similar with but different from the continuum waveguide, the coupled-resonator waveguide provides an alternative structure for manipulating the spatial and spectral properties of photons, where photon transport can be controlled by designing combinations of resonators with the nonlinearity of the resonator \cite{Marin-Palomo2017,Stern2018,Zhang2019,Lu2019,Szabados2020,Yu2021np} or by actively connecting resonators with dynamic modulations \cite{Fang2012,Fang2013,Williamson2020}.
In both cases, atoms (or quantum emitters) can be added into the coupled-resonator waveguide and hence further possible controllability of photons has been discussed \cite{Hoffman2011,Zhou2013,Fitzpatrick2017,Wang2020,Wang2021}.
Although some models are originally designed for photonic structures, such coupled-resonator waveguide has also been extensively discussed in the on-chip platform of superconducting transmission line resonators \cite{Devoret2007,GU2017,Blais2021}, where microwave photons transport and can be interacting with the artificial superconducting qubit \cite{Hoi2011,Stockklauser2017,Fitzpatrick2017,Yan2019}.

Recently, the atom-waveguide system has been generalized to studies of interactions between the photon in the waveguide and a \textit{giant} atom, where an artificial atom (quantum emitter) is fabricated to couple multiple points on the waveguide.
Due to the fact that multi-path quantum interferences are included in interactions between waveguide photons and giant atoms, a variety of interesting quantum optical phenomena have been explored, including bound states or dressed states \cite{Zhao2020,Guo2020,Wang2021,Cheng2021,Vega2021},  decoherence-free interaction \cite{Kockum2018,Kannan2020,Carollo2020,Soro2021}, electromagnetically-induced transparency \cite{Ask2020,Vadiraj2021,Zhao2021,Zhu2021}, and many others \cite{Kockum2014,Gonzalez-Tudela2019,Cilluffo2020,Longhi2020,Du2021b,Du2021,Yu2021,Cai2021,Wang2021arXiv,Du2021c}.
Relevant experiments have also been demonstrated that microwave photons or propagating phonons have been successfully coupled to an artificial gaint atom \cite{Andersson2019,Kannan2020,Vadiraj2021}.
Hence, explorations of different opportunities in seeking exotic manipulations of photons via quantum interferences from the photon-giant-atom interaction in the coupled resonator waveguide trigger further theoretical interests.

\begin{figure}
\includegraphics[width=12cm]{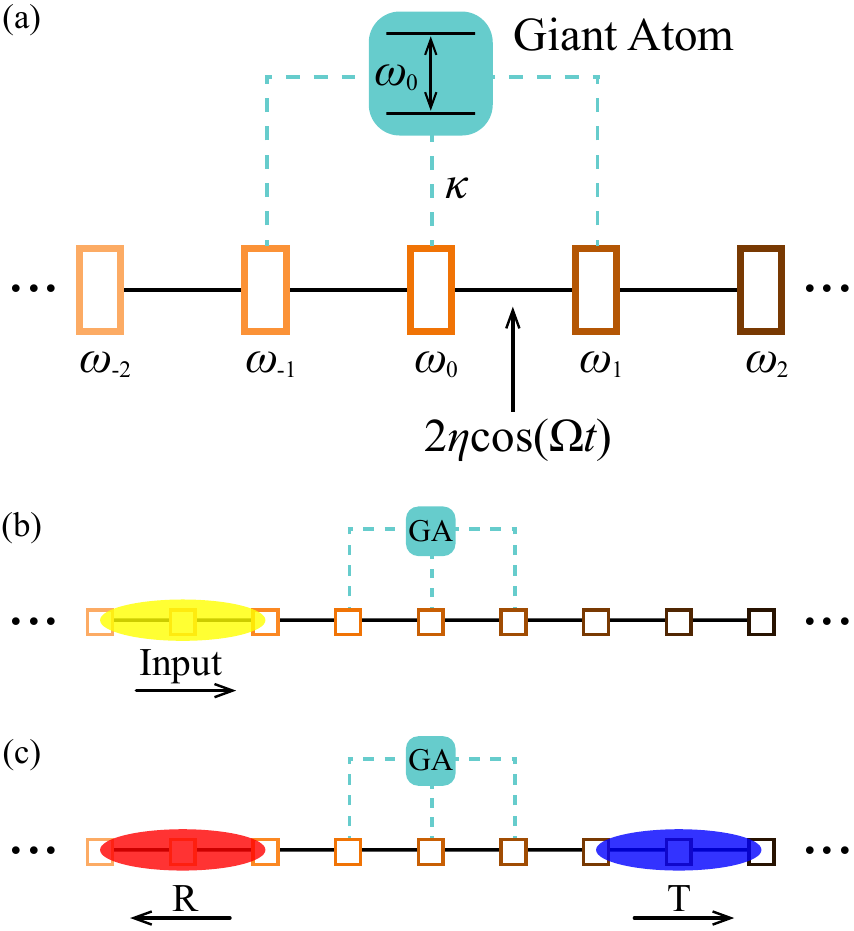}
\caption{(a) Schematics for a 1D dynamically modulated coupled-resonator waveguide coupling to a two-level giant atom. (b) The excited source is injected into the waveguide (yellow). After it interacting with the giant atom, the field is transmitted (blue) or reflected (red).}\label{fig1}
\end{figure}

In this paper, we study a theoretical model of an artificial two-level giant atom coupled with dynamically-modulated coupled-resonator waveguide [see Fig.~\ref{fig1}(a)], where each resonator (labelled by $m$) supports a resonant mode at the frequency $\omega_m=\omega_0+m\Omega$ with $\omega_0$ being the transition frequency of the atom.
The giant atom couples to the middle three resonators ($m=0,\pm 1$) through three separate paths.
We find that, for the specific choice of parameters, the wavepacket of the photon that transports inside the waveguide exhibits the localization effect, where the time of the photon-atom interaction lasts longer than the coupling rate between the atom and the resonator, pointing towards the efficient excitation of a \textit{bound} state, which is fundamentally different from the common sense that localized light is difficult to be excited by propagating photons.
We provide the analytical analysis based on the separation of \textit{propagating} states and \textit{localized} states of light, and find the quantum transition from propagating states to localized states.
Our work therefore provides a unique analytical method towards understanding quantum interferences between the giant atom and the coupled-resonator waveguide, and also points out important active manipulation of the photon-atom interaction, which shall finds potential applications in the quantum information processing \cite{Wendin2017,Blais2020}.

As schematically shown in Fig.~\ref{fig1}(a), we consider a one-dimensional photonic resonator lattice, with each resonator supporting a single resonance at $\omega_m$. 
The dynamic modulation can be applied in-between two adjacent resonators by modulating two resonances in the auxiliary resonator with a sinusoid external source $2 \eta {\rm cos} \Omega t$ where $\Omega \ll \omega_0$ is the frequency and $\eta$ is the modulation amplitude \cite{Fang2012}.
A two-level giant atom is designed to couple with the $0$-th and $\pm1$-st resonators with the coupling strength $\kappa$.
By assuming $\hbar = 1$, the corresponding Hamiltonian is
\begin{equation}
H = \omega_{0} \sigma_{z} /2 + \sum_{m} \omega_{m} a_{m}^{\dagger} a_{m} + \sum_{m} 2 \eta {\rm cos}(\Omega t)(a_{m}^{\dagger} a_{m+1} + a_{m+1}^{\dagger} a_{m}) + \sum_{m=-1,0,1} \kappa (a_{m}^{\dagger} \sigma_{-} + a_{m} \sigma_{+}). \label{eq1}
\end{equation}
Here, $\sigma_{z} = [\sigma_{+},\sigma_{-}]$.
$\sigma_{+} = \ket{e}\bra{g}$ ($\sigma_{-} = \ket{g}\bra{e}$) is the ladder operator that transits the atom from ground state $\ket{g}$ to excited state $\ket{e}$ (and vice versa), and $a_{m}^{\dagger}$ ($a_{m}$) is the creation (annihilation) operator for the photon in the $m$-th resonator.
One can re-write the Hamiltonian in the interaction picture under the rotating-wave approximation (RWA) \cite{Scully1997}
\begin{equation}
V(t) 
= \sum_{m} \eta \left(a_{m}^{\dagger}a_{m+1} + a_{m+1}^{\dagger}a_{m}\right) + \sum_{m=-1,0,1} \kappa \left(a_{m}^{\dagger}\sigma_{-}e^{im\Omega t} + a_{m}\sigma_{+}e^{-im\Omega t}\right). \label{eq2}
\end{equation}
To simulate the dynamics of photon transport, we write the single-excitation wave function
\begin{equation}
\ket{\psi(t)} = \sum_{m} v_{m}(t) a_{m}^{\dagger} \ket{0,g} + \xi(t) \ket{0,e}, \label{eq3}
\end{equation}
where $v_{m}$ is the probability amplitude for creating the photon from the vacuum state $\ket{0}$ in the $m$-th resonator while the atom remains at the ground state $\ket{g}$, and $\xi$ is the probability amplitude for the atom being excited (to $\ket{e}$) by the propagating photon.
By using the Schr\"{o}dinger's equation, we obtain the working equations for simulations
\begin{equation}
\dot{v}_{m} = -i \eta \left(v_{m+1} + v_{m-1}\right) - i\kappa \xi e^{im\Omega t} \delta_{m,m'}, ~ \dot{\xi} = -i \kappa \sum_{m=-1,0,1} v_{m}e^{-im\Omega t}, \label{eq4}
\end{equation}
where $m'$ denotes $0$ or $\pm 1$.

\begin{figure}
\includegraphics[width=16cm]{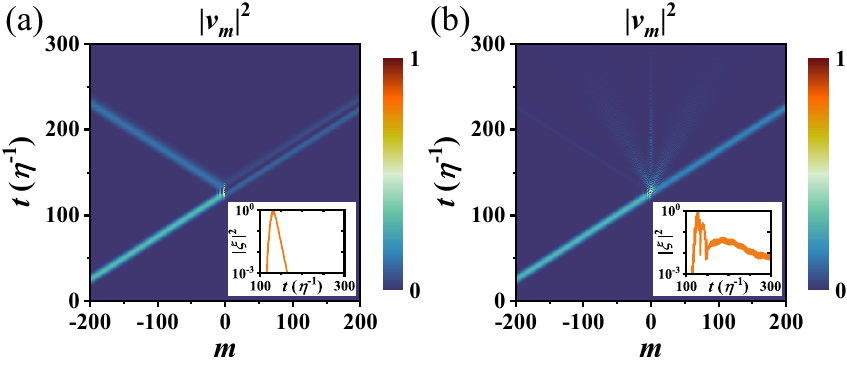}
\caption{The probability distribution of photon in different resonators versus the time, when (a) $\Omega=3 \eta$ and (b) $\Omega=2.05 \eta$, respectively. The insets show the corresponding atomic excitation probability in the logarithm scale.}\label{fig2}
\end{figure}

In simulations, we consider a coupled-resonator waveguide composed by $401$ resonators ($m = -200,\cdots,200$).
A Gaussian-shape pulse $S = e^{-(t-t_{0})^{2}/\tau^{2}}$ is used to excite the leftmost resonator, where $\tau=5\sqrt{2} \eta^{-1}$ and $t_{0} =25 \eta^{-1}$.
We first consider the case that $\kappa=0.5 \eta$ and $\Omega=3 \eta$. In Fig.~\ref{fig2}(a), we plot the distribution of $|v_m|^2$ on different resonators and $|\xi|^2$ versus the time $t$.
One sees that the photon is injected into the system from the left and then propagates towards the right. Once it interacts with the atom, a portion of the wavepacket of the photon is reflected while the atom is excited. The spectrum of the transmitted wavepacket is given in the supplementary, which shows consistence with the propagating photon interacting with a resonant atom in a waveguide \cite{Shen2005,Shen2007}.

The striking feature of the system is found when we choose $\Omega=2.05 \eta$ while keeping other parameters unchanged, with simulation results plotted in Fig.~\ref{fig2}(b).
One sees that, once the light interacts with the atom in the vicinity of the $0$-th resonator, the wavepacket of the photon is stored in the vicinity of the middle three resonators for a relative long time ($\sim 150 \eta^{-1} \gg \kappa^{-1}$) and the excitation of the atom also gives a relative long decay tail.
Such the phenomenon denotes a \textit{bound} state of photon and atom where the light shows the localization effect near middle resonators and atom exhibits a subradiant decay time.
Besides the bound state, together with the transmission of a portion of wavepacket at the original group velocity, other portions of wavepacket are transmitted and reflected at a smaller group velocity.

In order to understand the bound state in our proposed model analytically, we next analyze the Hamiltonian (\ref{eq2}) in details.
The first term in Eq.~(\ref{eq2}) gives interactions between resonators driven by an external source that composes the waveguide, while the second term describes atom-resonator interactions, where $\pm \Omega$ represents the detuning between the $\pm 1$-st resonator and the atom. 
Since the atom only couples with the middle three resonators, and we consider a finite number of resonators ($-d \leq m < d$ with $d$ being a positive integer), the influence of the photon state in resonators at two boundaries is negligible for $d \gg 1$. 
We hence can take the state $\ket{k} = \sum\limits_{m=-d}^{d-1} a_{m}^{\dagger} \ket{0} e^{i mk\pi /d} / \sqrt{2d}$ ($k = -d,\cdots,d-1$) in the momentum space which is the eigenstate of $\sum\limits_{m=-d}^{d-1} \eta (a_{m}^{\dagger}a_{m+1} + {\rm h.c.})$ and is regarded as the Bloch wave in the lattice with mode frequency $\omega_{k} = 2 \eta \cos(k\pi /d)$.
It also worth mentioning that the group velocity of the wavepacket is $v_{g} = -2 \eta \sin(k\pi /d)$.
Obviously, when $\omega_{k} = \pm 2\eta$, the corresponding wavepacket has $v_{g}=0$ and therefore does not move in the lattice. We will refer to them as localized modes in the following discussions.

Along with the atomic states $\ket{e}$ and $\ket{g}$, now we can rewrite $V$ in the $k$-space, which leads to $V = V_{0} + V_{1}$ where $V_{0} = \sum\limits_{k=-d}^{d-1} \omega_{k} \ket{k} \bra{k}$ and $V_{1} = \sum\limits_{k=-d}^{d-1} \sum\limits_{m=-1}^{1} \kappa (e^{-i mk\pi /d} e^{im \Omega t} \ket{k,g}\bra{0,e} + \rm{h.c.}) /\sqrt{2d}$.
If we consider the interaction-picture Hamiltonian under the $k$ representation, namely $e^{iV_{0}t} V_{1} e^{-iV_{0}t} = \sum\limits_{k=-d}^{d-1} \sum\limits_{m=-1}^{1} \kappa (e^{-i mk\pi /d} e^{i(m\Omega + \omega_{k})t} \ket{k,g}\bra{0,e} + \rm{h.c.}) /\sqrt{2d}$, one clearly sees that each $k$-mode interacts with the atom through three resonators $m = 0,\pm1$, while $m\Omega+\omega_{k} = 0$ represents the resonance condition between the atom and the photon state.
Following this argument, we can assume that each $k$-mode is only coupled with the resonator closest to satisfy the resonance condition.
Consequently, we obtain the Hamiltonian as
\begin{equation}
V \approx \sum_{k=-d}^{d-1} \omega_{k} \ket{k} \bra{k} + \frac{\kappa}{\sqrt{2d}} \sum_{m=-1}^{1} \left(\sum_{k \in K_{m}} e^{-i \frac{mk\pi}{d}} e^{im\Omega t} \ket{k,g}\bra{0,e} + \rm{h.c.} \right), \label{eq5} 
\end{equation}
where we divide $k$ into three regions.
In region $K_{0}$, the photon state of the Bloch wave has $\omega_{k} \in (-\sqrt{2}\eta,\sqrt{2}\eta)$, and is only coupled with the resonator $m = 0$.
Similarly, in region $K_{\pm 1}$, $\omega_{k} \in (\mp \sqrt{2}\eta,\mp 2\eta]$, and the photon state is coupled with the resonator $m = \pm1$.
We emphasize that, according to the resonance condition, the frequencies of the $k$-mode in the photon state that make the most contribution are $\omega_{k} = 0, \pm 2\eta$.
Therefore, except for the modes near these three frequencies, other $k$-modes are negligible in our analytical analysis, and the choice of the limit of the above-mentioned regions is for the convenience purpose.

Assuming that the wave function of the photon state has the form
\begin{equation}
\ket{\psi(t)}_{k} = \sum_{k=-d}^{d-1} C_{k}(t) \ket{k,g} + \chi(t) \ket{0,e}, \label{eq6}
\end{equation}
we can get the corresponding dynamic evolution equations from Eq.~(\ref{eq5}).
However, we notice that, there are $4$ corresponding states $\ket{k}$ at each $|\omega_{k}|$. 
For the sake of simplicity, we hence define $J_{k,s\pm}(t) = \{[C_{k}(t) e^{-i k\pi /d} + C_{-k}(t) e^{ik \pi /d}]e^{i\Omega t} \pm [C_{d-k}(t) e^{i (\pi-k\pi/d)} + C_{-(d-k)}(t) e^{-i(\pi-k\pi/d)}]e^{-i\Omega t}\}/2$ and $J_{k,0\pm}(t) = \{[C_{d/2-k}(t) + C_{-(d/2-k)}(t)] \pm [C_{d/2+k}(t) + C_{-(d/2+k)}(t)]\}/2$ with $k \in [0,d/4)$, and obtain
\begin{gather}
i\frac{\partial}{\partial t} J_{k,s+} (t) = \frac{2\kappa}{\sqrt{2d}}\chi(t) - (\Omega-\omega_{k}) J_{k,s-}(t), \label{eq7} \\
i\frac{\partial}{\partial t} J_{k,s-} (t) = - (\Omega-\omega_{k}) J_{k,s+}(t), \label{eq8} \\
i\frac{\partial}{\partial t} \chi(t) = \frac{\kappa}{\sqrt{2d}} \left[J_{0,s+}(t) + 2\sum_{k=1}^{d/4-1} J_{k,s+}(t)\right] + \frac{\kappa}{\sqrt{2d}} \left[J_{0,0+}(t) + 2\sum_{k=1}^{d/4-1} J_{k,0+}(t) \right], \label{eq9} \\
i\frac{\partial}{\partial t} J_{k,0+} (t) = \frac{2\kappa}{\sqrt{2d}}\chi(t) - v_{g} J_{k,0-}(t), \label{eq10} \\
i\frac{\partial}{\partial t} J_{k,0-} (t) = - v_{g} J_{k,0+}(t). \label{eq11}
\end{gather}
Here we emphasize that $J_{k,s\pm}$ denotes the modes of the photon state whose group velocities belong to $(\pm \sqrt{2} \eta,0]$ (where $J_{0,s\pm}$ corresponds to modes with $v_{g}=0$, i.e., \textit{localized} states), and $J_{k,0\pm}$ denotes those modes whose group velocities belong to $(\pm \sqrt{2} \eta, \pm 2\eta]$ (where $J_{0,0\pm}$ corresponds to modes with $v_{g}=\pm 2\eta$, i.e., \textit{propagating} states).
Eqs.~(\ref{eq7})-(\ref{eq11}) indicates that $J_{k,s\pm}$ and $J_{k,0\pm}$ form two sets of subsystems, connected by the atom [see Eq.~(\ref{eq9})].
Since the localized modes are included, let us focus on the subsystem described by Eqs.~(\ref{eq7})-(\ref{eq9}) then.
In order to erase the impact of the other subsystem, we discard the second term on the right-hand side of Eq.~(\ref{eq9}).
Then, by setting that $J_{k,s\pm}(t) = j_{k,s\pm}e^{-i\lambda_{s}t}$ and $\chi(t) = x_{s} e^{-i \lambda_{s} t}$, we find the special solution with $\lambda_{s}=0$, which gives
\begin{equation}
x_{s} = \frac{\sqrt{2d}}{2\kappa} (\Omega - \omega_{k}) j_{k,s-}, ~ j_{k,s+}=0. \label{eq12}
\end{equation}
Eq.~(\ref{eq12}) is solvable with the normalization condition involved.
Eq.~(\ref{eq12}) suggests that when $\Omega > \eta$, $|j_{k,s-}|$ decreases as $k$ increases, which implies that the mode distribution of the photon state is concentrated near $\omega_{k} = \pm 2\eta$.
Notice that $|j_{k,s-}|$ is proportional to $|x_{s}|$.
Once the atom is excited, it is possible to observe the localized state with $v_{g}=0$.
This is a bound state in which the photon is stored and the atom keeps excited with the subradiant decay.
However, we need to point out that when $\Omega$ becomes large, the proportion of the atomic excited state in the bound state is also increasing.
Due to normalization condition, the probability that the system is in the localized state in the vicinity of $k=0$ is suppressed.
On the other hand, when $\Omega < \eta$, $|j_{k,s-}|$ increases as $k$ increases, and hence the maximum value of $|j_{k,s-}|$ is at $k \neq 0$ ($v_{g} \neq 0$).
In such case, the localized state is nearly impossible to be generated.
Therefore, we only have a very narrow window of $\Omega$ to obtain a significant localized photon state, which is consistent with our previous numerical results.

The above discussion shows that through the interaction with the $\pm 1$-st resonators and the atom, the photon may move slowly or even stay localized in the waveguide.
The next question is how to excite such localized state, since the initial wavepacket of the photon is prepared centered at $\omega_{k}=0$ with spectral width $\sim 1/\tau$ nearly does not contain any component of $\omega_{k} = \pm 2\eta$ ($v_{g}=0$).
From Eq.~(\ref{eq9}), one finds that the coupling strengths of the two subsystems and the atom are similar, which means that the effects of the two subsystems on the atom are comparable.
Consequently, when the initial state of the photon is prepared in one subsystem (corresponding to $J_{k,0\pm}$) where the center frequency of the initial wavepacket is $\omega_{k}=0$ here, this causes the excitation of the atom and in turn causes the excitation of another subsystem (corresponding to $J_{p,s\pm}$), i.e., creating the bound state with the localized photon state and excited atom at a subradiant decay.

\begin{figure}
\includegraphics[width=16cm]{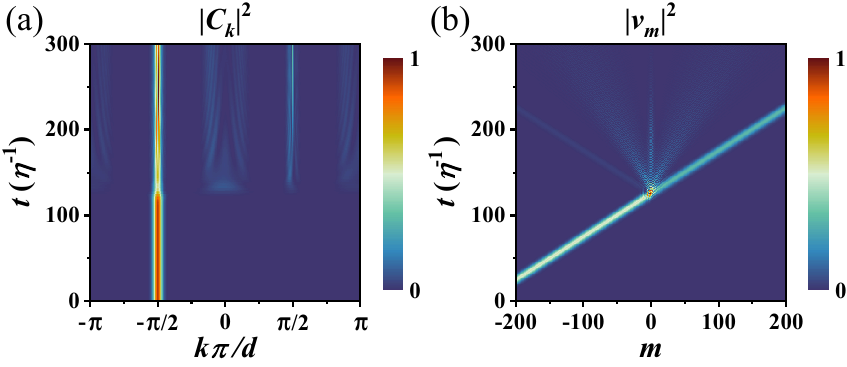}
\caption{(a) The evolution of the photon's wavepacket in the $k$-space. (b) The probability distribution of photon in different resonators versus the time. In both figures, simulations are performed in the $k$-space with $\Omega=2.05 \eta$.}\label{fig3}
\end{figure}

Next, we perform simulations in the $k$-space with the separation approximation that we analytically discussed above.
By combining Eq.~(\ref{eq5}) and the Schr\"{o}dinger's equation, we simulate the evolution of the states in the $k$-space over time with $\Omega=2.05\eta$ and plot the result in Fig.~\ref{fig3}(a).
The initial state of the system is assumed to be a Gaussian wavepacket in real space, with its center position at $m_{0} = -250$ and propagating toward the right at a group velocity $2\eta$, i.e.
$\ket{\psi_{I}} = A_{0}\sum\limits_{m=-d}^{d}e^{-(m-m_{0})^2/2\delta_{m}^2}e^{-i\pi (m-m_{0})/2}\ket{m}$, where $A_{0}$ is the normalization coefficient and $\delta_{m} = 10$.
(The choice of the initial wavepacket is consistent with the boundary-excitation source in simulations for Fig.~\ref{fig2}.)
Therefore, the Fourier transformation of the initial state (with taking $d = 500$) here gives the initial wave function in Eq.~(\ref{eq6}) for simulations, with the central momentum being $k\pi/d = -\pi/2$. 
Fig.~\ref{fig3}(a) indicates that the mode separation approximation we used in the above discussion is feasible, because one can see that the energy distribution of the photon's wavepacket is centered near $k\pi/d = -\pi,-\pi/2,0,\pi/2$ with clear separations.
Next, we follow the mode separation approximation and use Eqs.~(\ref{eq7})-(\ref{eq11}) with the same parameters for Fig.~\ref{fig3}(a) and numerically calculate the photon distribution in the $k$-space.
After we Fourier-transform the simulation results back into the real space, we plot the resulting probability distribution of photon in Fig.~\ref{fig3}(b).
One can see that evolutions of the photon in both Fig.~\ref{fig3}(b) and Fig.~\ref{fig2}(b) match quite well, indicating our analytical analysis is useful in understanding such light localization phenomenon.

\begin{figure}
\includegraphics[width=12cm]{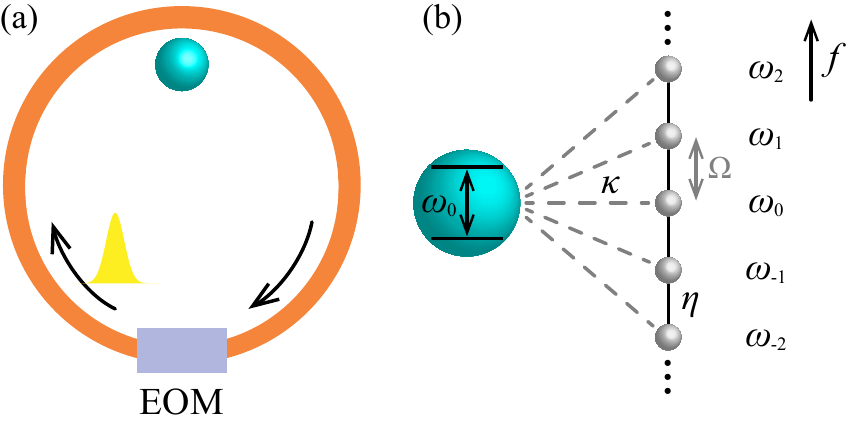}
\caption{(a) Schematic for a dynamically modulated ring resonator coupling to a two-level giant atom. EOM denotes to the electro-optic modulator. (b) The two-level atom coupled with the lattice in the synthetic frequency dimension.}\label{fig4}
\end{figure}

Before we conclude our paper, we notice that our studied model naturally supports a synthetic lattice along the frequency axis of light, which is constructed in a modulated ring resonator and is similarly described by the Hamiltonian $\sum\limits_{m} \omega_{m} a_{m}^{\dagger} a_{m} + \sum\limits_{m} 2 \eta {\rm cos}(\Omega t)(a_{m}^{\dagger} a_{m+1} + a_{m+1}^{\dagger} a_{m})$ (the same as with the waveguide part in Eq.~(\ref{eq1})) with $\omega_{m}$ being resonant frequencies and $\Omega$ is the frequency of the modulator \cite{Yuan2021}.
Once a two-level atom at the transition frequency $\omega_{0}$ is added to couple with the ring, an effective giant atom coupled with the synthetic lattice with the linear-gradient detuning ($\omega_{m}-\omega_{0}$) at each connection is built, as shown in Fig.~\ref{fig4}.
We find that the affects from far-from-resonance couplings weakly affect the system, and our findings in this work shall also map to the dynamics in the synthetic dimension (see the supplementary for details). However, detailed consequences in the output of the single-photon spectrum desires further studies with the input-output formalism carefully included, which is beyond the scope of this paper. Yet, our work is still useful in the future understanding of the photon-transport problem in a synthetic frequency dimension coupled with a two-level atom. 

In summary, we study the photon propagation problem inside a coupled-resonator waveguide under the dynamic modulation with the middle three resonators coupled with the giant atom, where the dynamic modulation frequency precisely equals to the frequency difference between two nearby resonators.
We find that, through multi-resoantor couplings, one can efficiently excite the static Bloch wave mode in the system by a propagating input photon, so the light field can be localized for a long time and the atom exhibits subradiant decay.
Our model is valid for a variety of potential experimental platforms, including photonic-crystal waveguide \cite{Fang2012}, coupled cavities in free space \cite{Peng2014,Jiang2016}, and superconducting transmission line resonators \cite{Wallraff2004,Yin2013,Pechal2014}.
Our work therefore shows a theoretical perspective for studying photon-atom interactions in waveguide systems and seeks additional external control of the propagating photon, which is fundamentally important for the quantum manipulation of a single photon.

The research is supported by National Natural Science Foundation of China (12122407, 11704241, and 11974245), National Key R$\&$D Program of China (2017YFA0303701), Shanghai Municipal Science and Technology Major Project (2019SHZDZX01), and Natural Science Foundation of Shanghai (19ZR1475700). L.Y. acknowledges the support from the Program for Professor of Special Appointment (Eastern Scholar) at Shanghai Institutions of Higher Learning. X.C. also acknowledges the support from Shandong Quancheng Scholarship (00242019024).




%


\end{document}